\documentclass[3p,times]{elsarticle}

\usepackage{ecrc}

\volume{00}

\firstpage{1}

\journalname{Procedia Computer Science}

\runauth{}

\jid{procs}

\jnltitlelogo{Procedia Computer Science}

\CopyrightLine{2011}{Published by Elsevier Ltd.}

\usepackage{amssymb}

\usepackage[figuresright]{rotating}
\usepackage{graphicx}

\begin{document}

\begin{frontmatter}



\dochead{}

\title{
Performance Evaluation of Delay Tolerant Network in Heterogeneous Highly Dense Mobile Environment 
}
\author[label1]{R.S.Mangrulkar}
\author[label2]{Dr. Mohammad Atique}
\fntext[label1]{{Research Scholar, SGBAU Amravati, Maharashtra, India.  Associate Professor, Bapurao Deshmukh College of Engineering,Sevagram, Wardha, Maharashtra, India} \fnref{label1}}
\ead[label1]{rsmangrulkar@gmail.com} 
\fntext[label2]{{Associate Professor, Post Graduate Department of Computer Science, S.G.B.A.U, Amravati, Maharashtra, India} \fnref{label2}}
\ead[label2]{mohd.atique@gmail.com}



\begin{abstract}
Delay tolerant network (DTN) is opportunistic network where each node searches best opportunity to deliver the message called bundle to the destination. DTN implements a store and forward message switching system by simply introducing another new protocol layer called the Bundle Layer on top of the transport layer. The bundle layer is responsible for storing and forwarding entire message in message segments called bundles between source node and destination node. This paper evaluates the performance of delay tolerant network layer in heterogeneous highly dense mobile node environment. The heterogeneous network is created with the help of stationary wired node and Base Station node by introducing dynamic dense Mobile node network. Mobile nodes are assigned with continuous mobility. Three parameters are suggested $\Delta$, $\Theta$ and $\lambda$ to correlate the results obtained using rigorous simulation. Results show that after some threshold values, dense feature about mobile node does not pretend the delay cause for delay tolerant network packets. Also, increase in number of mobile node and number of File Transfer connection rarely change the overall performance of the delay tolerant network.
\end{abstract}


\end{frontmatter}


\section{Introduction}\label{intro}
Delay tolerant network (DTN) is opportunistic network where every node searches best opportunity to deliver the message called bundle to the destination. Mobile ad hoc network supposed to be the origin of delay tolerant network. In traditional Manet, node called source requires end-to-end connection to send data. This almost produces the high possibility of data transmission even in short period of time. In DTN, end-to-end connection is never ensured between source node and destination node which involves large delay for data sent. Node called source always search the best opportunity to pass on message to the destination through some sympathetic and trustworthy intermediate node. Support and trust is wished as source has to believe on intermediate nodes. Therefore delay tolerant network transmission is also called as cooperative transmission.
Conventional TCP/IP model is unsuitable for delay tolerant network application. In conventional model end-to-end connectivity is needed from source to destination to send data. Perhaps, TCP is responsible for error free transmission because it has reliable connection oriented nature. But in DTN, there is no path selected as fixed path for transmission. Node called source has to shift message to the nearest intermediate node towards destination and wait for acknoledgment. 
This failure of conventional TCP/IP model demands architecture variation in existing TCP/IP model. For this, Bundle layer is introduced in between application layer and transport layer. Source node creates message called bundle and hand over to the immediate next layer called bundle layer. This bundle layer argue track on each bundle sent in the network. Even though, path is not fixed and reserved from source to destination, cooperative nature of nodes help bundle to be carried to the destination. The Bundle layer node keeps track on the number of bundle created sent successfully and lost if at all. This tracking helps source node to estimate the cooperation of the intermediate nodes involved in bundle transmission. Intermediate node sends back the custodian report called acknoledgment to the source node or the immediate back node in the path towards destination. In this way message called bundle sent with the cooperation of all the nodes in the network. In this paper we carried out implementation and simulation of delay tolerant network in a heterogeneous high density mobile node environment. For subsequent scenarios considered, Mobile node density is increased to note the influence on delay tolerant network packet sent in the network. With increasing Mobile node density, we increase the delay tolerant network traffic by increasing the number of FTP connections setup between stationary \emph{Wired node} to \emph{Mobile node} using intermediate gateway \emph{Base station node}. 
Rest of the section is organized as: section \ref{rel} gives the related work found in the literature on Delay Tolerant Network on some open issues.  Section \ref{des} gives the brief design of delay tolerant network. Section \ref{sim} gives the details of simulations we carried for all scenario. The Results and their analysis are given in Section \ref{res} followed by conclusion in section \ref{conl}. 
\section{Related work} \label{rel}
~\cite{nr1} exposes rarely solved and highly emerging issue that is Data streaming over Delay Tolerant Networks. Because of demanding nature of streaming applications and their wide applications, this issue of data streaming is to be solved. Authors proposed Bundle Streaming Service (BSS) as a framework to improve the reception and storage of data streams. Authors have designed BSS by considering Interplanetary Internet (IPN) and its associated issues. They mention the open issue of data streaming such as out-of-order delivered packet ratio, packet loss ratio, frame loss ratio and peak signal-to-noise ratio, in order to accurately assess the efficacy of BSS and evaluate the impact of frame size, hop count and mobility on its performance. Thus the mobility issue and its associated experiments are yet to be exposed ~\cite {nr3}.We pick this issue and exposed with heterogeneous highly dense environment of Mobile nodes. \\
~\cite{nr2} mentioned use of erasure coding techniques in order to construct a disruption tolerant video sequence so that in the event of disruption, helpful video content is still provided to clients by injecting additional "summary frames" to the original stream ~\cite{nr4}.\\
~\cite{nr6}introduce a scheme where Home Access Points are equipped with storage capabilities and offer connectivity to mobile users. Whenever connectivity sharing is unavailable, the access points store the pending data and transmit it
as soon as connectivity becomes available. Experiment with realistic scenarios based on real network configurations and mobility maps are explains ~\cite {nr6}.\\
In ~\cite{nr7} the benefits of erasure coding for file transfers over space is explored using a generic end-to-end mechanism built on top of
The Bundle Protocol that incorporates LDPC codes. It also highlights the ability of erasure coding to provide different QoS to applications, in terms of file delivery latency, by properly tuning the code rate. \\
In ~\cite{nr8} advantages and disadvantages of having a transport protocol relying on DTN for providing reliable and transparent transfer of data between end systems and offering common transport layer services such as end-to-end
Packet-oriented retransmission, flow control, path diversity and redundant transmission. In that context some features from DS-TP and DTTP that correspond to the requirements of transport protocols for space are highlighted.\\
In ~\cite{nr9} specify the suitable architecture of Delay tolerant network in static and stationary environment. In this they define the steps to be carried out to implement Delay Tolerant Network model in network simulator NS2. Also they throw some light on open issues as to implement the DTN model and test is the highly dense mobile environment. We took this issue as challenge and provide some simulation flavours to solve this issue and carried some experimentation work described in section \ref{sim} using NS2 in highly dense mobile nodes to discuss and solve this issue.  \\
\section{Design of Delay Tolerant Network}\label{des}
\begin{figure}[h]
\centering
\mbox{\includegraphics[width=3in, height=2in]{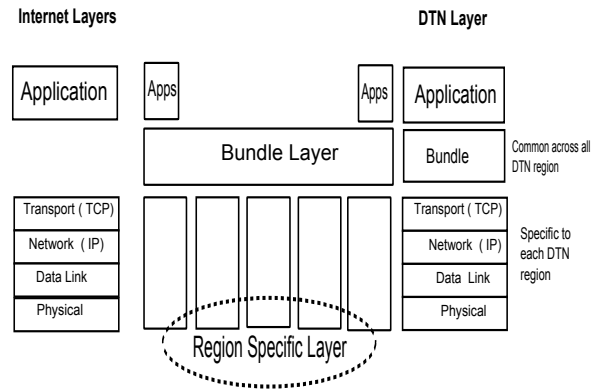}}
\caption{Layer Communication in DTN}
\label{f1}
\end{figure}

\begin{figure}[h]
\centering
\mbox{\includegraphics[width=3in, height=2in]{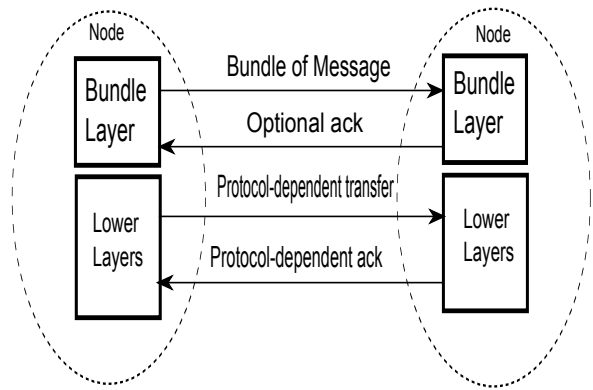}}
\caption{Node Communication in DTN}
\label{f2}
\end{figure}

\begin{figure}[h]
\centering
\mbox{\includegraphics[width=3in,height=2in]{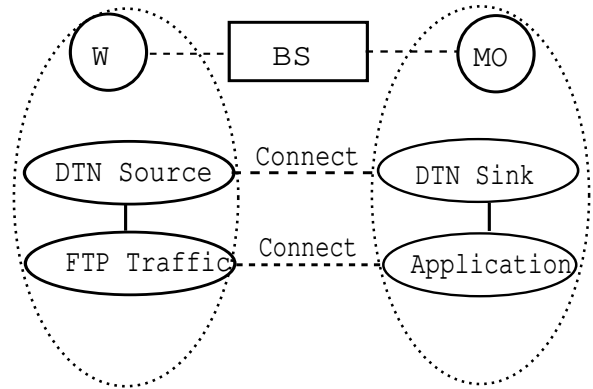}}
\caption{Node Configuration while Communication in DTN}
\label{f3}
\end{figure}

Delay Tolerant Network uses a store and forward message switching by simply introducing another new protocol layer called the Bundle Layer on top of the transport layer. The bundle layer responsible for storing and forwarding entire bundles between source node and destination node. Nodes interested to send data must have the bundle layer support on top of TCP layer.  The transport layer protocol either TCP or UDP below the bundle layer is selected based upon the reachability and accessibility properties estimated about each region in and around the intended node in the network. The figure \ref{f1} shows the bundle layer and its position in the existing TCP/IP internet protocol suit. As TCP protocol is responsible for communication between two end, it is called \lq\lq end-to-end communication\rq\rq , whereas Network layer perform \lq\lq host to host communication\rq\rq. In the similar way, DTN layer at node called source communicates with another node called destination using unit of transmission as bundle. Therefore the communication if called \lq\lq Bundle-to-Bundle communication\rq\rq. 
Figure \ref{f2} suggests the layer-to-layer communication in Delay Tolerant Network where bundle of message or simply bundle is sent from source node to target destination. Acknowledging depends on the transport layer protocol, either UDP or TCP and therefore it is optional. Lower layer communicates using underlying network protocols. So, it is called protocol dependent transfer with optional acknowledgment.
Figure \ref{f3} shows the communication model considered for Delay Tolerant Network implementation and simulation in highly dense mobile environment. Wired node \emph {W} is connected with Mobile node \emph {M} by using intermediate Base station node \emph {BS} as gateway. Traffic towards mobile node is routed through base station node. Intermediate node, on receiving bundle, sends custodian report called acknowledgement to the source. Custodian report is sent by every intermediate node to the immediate previous node about receiving of bundle. Finally, destination node sends acknowledgment of bundles received to the last but one node as custodian acknowledgement.

\section{Simulation} \label{sim}
\subsection{Simulation Setup} \label{simsetup}
\begin{figure}[h]
\centering
\mbox{\includegraphics[width=3in, height=2in]{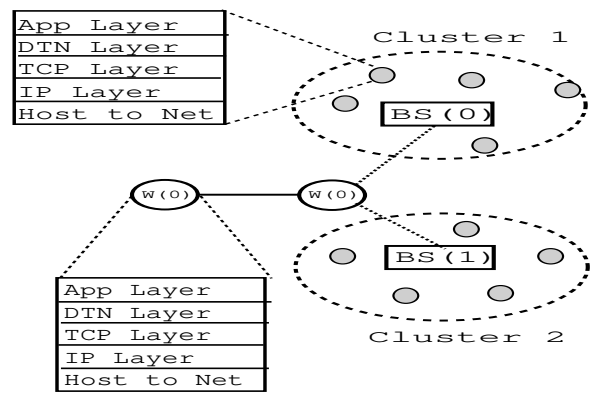}}
\caption{Network Topology common to all Scenario}
\label{f4}
\end{figure}

This section deals with the performance of delay tolerant network in heterogeneous environment with highly dense Mobile nodes. Before selecting what combination of number of nodes and File Transfer Protocol connection is best suited for our later study and development of new routing protocol for delay tolerant network, we need to see how delay tolerant network layer behaves in a heterogeneous mobile environment with variation in number of mobile node. The topology considered for all the scenario is given in figure \ref{f4}. Two clusters cluster \emph{1} and cluster \emph{2} are created. Six different scenario were considered by varying \emph{3} factors, \emph{1} Number of Mobile node, \emph {2} Number of FTP connection and \emph{3} destination node of File Transfer Protocol connection for simulating transmission of bundles in delay tolerant network. Source is always either Wired Node\emph{0} or Wired Node\emph{1}. Mobility scenario were created using \emph{setdest} command in \emph{NS2} under \emph{cmu-gen} utility. One group of Mobile Nodes created from half of the total number of mobile node is attached to Base Station \emph{0} and another half is attached to Base Station \emph{1}. Figure \ref{f4} shows the network topology considered for simulating subsequent scenarios \emph{1}. Traffic towards Mobile node \emph{M} is routed through Base station node \emph{0} or \emph{1} functioning as gateway node. Intermediate node, on receiving bundle, sends custodian message called acknowledgement to the source. After getting best opportunity, intermediate node reinitiates data transmission. The process of searching best opportunity is continuously repeated at every intermediate node and custodian acknoledgment is sent back to the node sending data packets. Finally, destination node sends acknowledgment of bundles received. 
\subsection{Simulation Parameters}\label{simpara}
Simulation parameters are precisely written in following tables. These parameters are divided into two parts. Part I given in table \ref{t1} introduces simulation parameters which are kept constant for all subsequent simulations for the entire scenario considered in this section \ref{sim}. Part II given in table \ref{t2} introduce variable parameters whose value changes according to the scenario under consideration.

\begin{table}[t]
\centering
\caption{Constant Parameters for Simulations}
\begin{tabular}{|l|l|l|}
\hline
Sr. No. & Parameter & Value \\
\hline
1 & Channel & WirelessChannel \\
2 & Propagation & TwoRayGround \\
3 & Physical Interface & WirelessPhy \\
4 & Interface Queue & Queue/DropTail/PriQueue \\
5 & Interface Queue Length & 100 Packets \\
6 & Adhoc Routing Protocol & SRPDTN \\
7 & Antena & OmniAntenna \\
8 & Topography & 800 Meter by 800 Meter\\
9 & Total Simulation Time & 100 Seconds\\
10 & Wired Nodes & 2 \\
11 & Base Station Nodes & 2 \\
12 & IEEE Model& 802.11 \\
13 & Data Rate & 11 Mega Bytes (MB)\\
14 & Basic Rate & 1MB \\
15 & TCP Window Size & 100 \\
16 & Size of Data to be transmitted & 10 Mega Bytes (MB) \\
17 & Size of TCP Segment & 1460 Bytes \\
\hline
\end{tabular}
\label{t1}
\end{table}

All above parameters were kept constant while varying the parameters given in table \ref {t2}. These parameters are separately mentioned in the scenario considered for subsequent simulation in subsections.
\begin{table}[b]
\centering
\caption{Variable Parameters for Simulations}
\begin{tabular}{|c|l|}
\hline
Sr. No. & Parameter \\
\hline
1 & Number of Mobile Nodes \\
2 & Number of FTP Connection \\
3 & Source and Destination pair \\
4 & Start time and End time of FTP \\
\hline
\end{tabular}
\label{t2}
\end{table}
As number of nodes considered is variable parameter as \emph{nn}, File Transfer Protocol agents are created using \emph{for loop} in TCL script written for simulation using Network Simulator \emph{2}. These agents called Delay tolerant network agents are connected to DTNSource agents as Wired Node \emph{0} or Wired Node \emph{1}. The loop control variable \emph{i} is used to control the connection between DTNSource agent and DTNSink agent. Loop used is \\
$$\emph{for [set i 0] ; [$i \le nn/10$] ; [i = i+2]}$$ connect \emph{i} in  cluster \emph{1} to Base Station \emph{0} \\
Whereas
$$\emph{for [set i nn/2] ; [$i \le nn*3/4$] ;[i= i+k]}$$ connect \emph{i} in  cluster \emph{2} to Base Station \emph{1} \\
where  $$k= 2^n  \hspace{1cm}  k = 1,2, \cdots , n $$ where $n$ is Scenario considered varies from  Scenario \emph{1} , Scenario  \emph{2} , $\cdots$  , Scenario  \emph{n}. Each \emph{n} pointing to scenario number 
\emph{1} for  \emph{10} mobile nodes , Scenario number  \emph{2} for \emph{30} mobile nodes , Scenario number \emph{3} for \emph{50} mobile nodes , Scenario number \emph{4} for \emph{100} mobile nodes, Scenario number \emph{5} for \emph{150} mobile nodes and Scenario number \emph{6} for \emph{200} mobile nodes. \\
Total number of nodes are grouped in to \emph{2} groups. First group is attached with Base Station \emph{0} in cluster \emph{1}. The File Transfer Protocol traffic is attached with the nodes with parameter \emph{i} as FTP(i) with start time \emph{10s} and end time \emph{80s} if $$i \le \frac{nn}{2}$$ On the other hand, FTP(i) is attached with Base Station node \emph{1} in cluster \emph{2} with start time \emph{20s} and end time \emph{90s} if $$i \ge \frac{nn}{2}$$ Following subsections clearly explain the scenario considered for simulations.
\subsection{Scenario 1: 10 Mobile Nodes}
For the simulation of Scenario \emph{1}, the variable parameters are given in table \ref {t3}. Number of Wired Nodes considered are \emph{2}, Number of Mobile Nodes considered are \emph{10}, Base Station nodes considered are \emph{2}. Half of the mobile nodes from Mobile Node \emph{(0)} to Mobile Node \emph{(4)} are connected to Base Station \emph{0} by considering cluster \emph{1} as a transmission area of Base Station \emph{0}. On the other hand, remaining half of the mobile nodes from Mobile node \emph{(5)} to Mobile Node \emph{(9)} are connected to  Base Station \emph{1} by considering cluster \emph{2} as transmission area of Base Station \emph{1}. The source - destination pair with their individual start time and end time of FTP in simulations are given in table \ref {t3}.
\begin{table}[t]
\centering
\caption{Parameters for Scenario 1}
\begin{tabular}{|c|c|c|}
\hline
Sr. NO. & Parameter & Value \\
\hline
1 & Number of Node & 10 \\
2 & Total FTP Connection & 02 \\
\hline
3 & \multicolumn{2}{|c|}{ FTP(0) in Cluster 1} \\
 & Start time: 10s &  End time:80s\\
 & Source Node: Wired Node(0) & Dest Node: Mobile Node(0) \\
\hline
4 & \multicolumn{2}{|c|}{ FTP(5) in Cluster 2} \\
 & Start time: 10s &  End time:90s\\
 & Source Node: Wired Node(0) & Dest Node: Mobile Node(5) \\
\hline
\end{tabular}
\label{t3}
\end{table}

File Transfer Protocol application is considered on top of Delay Tolerant Network layer designed in section \ref{des}. Delay tolerant network layer acts as intermediate layer called Bundle layer between file transfer as application layer and transport layer of protocol suit. The scenario is simulated in Network Simulator \emph{2}. Customization of delay tolerant network layer agent is done with modification in existing Network Simulator \emph{2.35} version which is used with File Transfer Protocol traffic using DTN agent.  The tracefile produced is analyzed for agent trace packets called DTN packets, Transmission control layer packets called (TCP), Network layer packets (RTR) and Mac layer Packets (MAC). The detail analysis and results are summarized in table \ref {t4}. For this we considered pause time of \emph{10s}. We developed awk script for analyzing the tracefile in one stroke and got the figures mention in table\ref{t4}.
\begin{table}[h]
\centering
\caption{Simulation Results of Scenario 1: \emph{10} Mobile Nodes}
\begin{tabular}{|c|c|c|c|c|c|}
\hline
Pause  & \multicolumn{5}{c|}{Packets} \\
\hline
Time(s) & DTN   & Overall & Received & Send & Drop \\
\hline
10 & 0 & 602103 & 200514 & 12 &  141 \\
20 & 2894 & 5934 & 2964 & 2925 & 28 \\
30 & 6943 & 13950 & 6970 & 6960 & 13 \\
40 & 8176 & 16410 & 8200 & 8188 & 16 \\
50 & 7792 & 15668 & 7840 & 7803 &  14 \\
60 & 7818 & 15714 & 7855 & 7833 &  18 \\
70 & 7914 & 15849 & 7921 & 7917 &  7 \\
80 & 7991 & 15984 & 7992 & 7991 &  1 \\
90 & 8116 & 16231 & 8115 & 8116 &  0 \\
100 & 7901 & 15815 & 7910 & 7903 &  1 \\
\hline
\end{tabular}
\label{t4}
\end{table}

\subsection{Scenario 2: 30 Mobile Nodes}
For this simulation of scenario 2, the variable parameters are given in table \ref {t5}. Number of Wired Nodes considered are \emph{2}, Number of Mobile Nodes considered are \emph{30}, Base Station nodes considered are \emph{2}. Half of the mobile nodes from Mobile Node \emph{(0)} to Mobile Node \emph{(14)} are connected to Base Station \emph{0} by considering cluster \emph{1} as a transmission area of Base Station \emph{0}. On the other hand, remaining half of the mobile nodes from Mobile node \emph{(15)} to Mobile Node \emph{(29)} are connected to  Base Station \emph{1} by considering cluster \emph{2} as transmission area of Base Station \emph{1}. The source - destination pair with their start time and end time of simulations are also given in table \ref {t5}.
\begin{table}[t]
\centering
\caption{Parameters for Scenario 2}
\begin{tabular}{|c|c|c|}
\hline
Sr. NO. & Parameter & Value \\
\hline
1 & Number of Node & 30 \\
2& Total FTP Connection & 05 \\
\hline
3 & \multicolumn{2}{|c|}{ FTP(0) in Cluster 1} \\
 & Start time: 10s &  End time:80s\\
 & Source Node: Wired Node(0) & Dest Node: Mobile Node(0) \\
\hline
4 & \multicolumn{2}{|c|}{ FTP(2) in Cluster 1} \\
 & Start time: 10s &  End time:90s\\
 & Source Node: Wired Node(0) & Dest Node: Mobile Node(2) \\
\hline
5 & \multicolumn{2}{|c|}{ FTP(15) in Cluster 2} \\
 & Start time: 20s &  End time:90s\\
 & Source Node: Wired Node(1) & Dest Node: Mobile Node(15) \\
\hline
$ \vdots $ & $ \cdots $ & $ \cdots $ \\
\hline
7 & \multicolumn{2}{|c|}{ FTP(21) in Cluster 2} \\
 & Start time: 20s &  End time:90s\\
 & Source Node: Wired Node(1) & Dest Node: Mobile Node(21) \\
\hline
\end{tabular}
\label{t5}
\end{table}
The details about the results with different types of packets and their agreeing numbers are summarized in table \ref {t6}.

\begin{table}[b]
\centering
\label{t6}
\caption{Simulation Results of Scenario 2:\emph{30} Mobile Nodes}
\begin{tabular}{|c|c|c|c|c|c|}
\hline
Pause  & \multicolumn{5}{c|}{Packets} \\
\hline
Time(s) & DTN   & Overall & Received & Send & Drop \\
\hline
10 & 0 & 612089 & 203822 & 12 &  196 \\
20 & 3415 & 7143 & 3611 & 3465 & 35\\
30 & 7051 & 14234 & 7132 & 7075 & 17 \\
40 & 8004 & 16058 & 8016 & 8015 & 25 \\
50 & 8087 & 16477 & 8311 & 8110 &  28 \\
60 & 7992 & 16117 & 8066 & 8010 &  31 \\
70 & 7875 & 15777 & 7893 & 7879 &  2\\
80 & 7909 & 15835 & 7921 & 7911 &  2 \\
90 & 8020 & 16044 & 8020 & 8021 &  3 \\
100 & 8013 & 16070 & 8049 & 8016 &  2 \\
\hline
\end{tabular}
\end{table}

\subsection{Scenario 3, 4, 5 and 6: 50,100,150 \& 200 Mobile Nodes}
For the subsequent scenario, the parameters which are considered are summarized in table ~\ref{t7},~\ref{t9} and ~\ref{t11}. Details about results with different types of packets and their corresponding numbers are summarized in table  ~\ref{t8},~\ref{t9} and ~\ref{t12}.
\begin{table}[t]
\centering
\caption{Parameters for Scenario 3}
\label{t7}
\begin{tabular}{|c|c|c|}
\hline
Sr. NO. & Parameter & Value \\
\hline
1 & Number of Node & 50 \\
2& Total FTP Connection & 07 \\
\hline
3 & \multicolumn{2}{|c|}{ FTP(0) in Cluster 1} \\
 & Start time: 10s &  End time:80s\\
 & Source Node: Wired Node(0) & Dest Node: Mobile Node(0) \\
\hline
$ \vdots $ & $ \cdots $ & $ \cdots $ \\
\hline
5 & \multicolumn{2}{|c|}{ FTP(4) in Cluster 1} \\
 & Start time: 10s &  End time:90s\\
 & Source Node: Wired Node(0) & Dest Node: Mobile Node(4) \\
\hline
6 & \multicolumn{2}{|c|}{ FTP(5) in Cluster 2} \\
 & Start time: 20s &  End time:90s\\
 & Source Node: Wired Node(1) & Dest Node: Mobile Node(5) \\
\hline
$ \vdots $ & $ \cdots $ & $ \cdots $ \\
\hline
9 & \multicolumn{2}{|c|}{ FTP(29) in Cluster 2} \\
 & Start time: 20s &  End time:90s\\
 & Source Node: Wired Node(1) & Dest Node: Mobile Node(29) \\
\hline
\end{tabular}
\end{table}

\begin{table}[t]
\centering
\caption{Simulation Results of Scenario 3:\emph{50} Mobile Nodes}
\label{t8}
\begin{tabular}{|c|c|c|c|c|c|}
\hline
Pause  & \multicolumn{5}{c|}{Packets} \\
\hline
Time(s) & DTN   & Overall & Received & Send & Drop \\
\hline
10 & 0 & 602679 & 200564 & 12 &  388 \\
20 & 2912 & 6211 & 3211 & 2951 & 26\\
30 & 7839 & 15816 & 7937 & 7858 & 14 \\
40 & 7711 & 15456 & 7719 & 7719 & 16 \\
50 & 7899 & 16168 & 8215 & 7915 &  19 \\
60 & 7820 & 15789 & 7927 & 7833 &  22 \\
70 & 8169 & 16363 & 8189 & 8171 &  1 \\
80 & 7892 & 15831 & 7930 & 7896 &  3 \\
90 & 7942 & 15890 & 7944 & 7943 &  3 \\
100 & 7920 & 15881 & 7956 & 7922 &  1 \\
\hline
\end{tabular}
\end{table}

\begin{table}[t]
\centering
\caption{Parameters for Scenario 4}
\label{t9}
\begin{tabular}{|c|c|c|}
\hline
Sr. NO. & Parameter & Value \\
\hline
1 & Number of Node & 100 \\
2 & Total FTP Connection & 09 \\
\hline
3 & \multicolumn{2}{|c|}{ FTP(0) in Cluster 1} \\
 & Start time: 10s &  End time:80s\\
 & Source Node: Wired Node(0) & Dest Node: Mobile Node(0) \\
$ \vdots $ & $ \cdots $ & $ \cdots $ \\
\hline
7 & \multicolumn{2}{|c|}{ FTP(8) in Cluster 1} \\
 & Start time: 10s &  End time:90s\\
 & Source Node: Wired Node(0) & Dest Node: Mobile Node(8) \\
\hline
8 & \multicolumn{2}{|c|}{ FTP(20) in Cluster 2} \\
 & Start time: 20s &  End time:90s\\
 & Source Node: Wired Node(1) & Dest Node: Mobile Node(20) \\
\hline
$ \vdots $ & $ \cdots $ & $ \cdots $ \\
\hline
11 & \multicolumn{2}{|c|}{ FTP(68) in Cluster 2} \\
 & Start time: 20s &  End time:90s\\
 & Source Node: Wired Node(1) & Dest Node: Mobile Node(68) \\
\hline
\end{tabular}
\end{table}

\begin{table}[t!]
\centering
\caption{Simulation Results of Scenario 4: \emph{100} Mobile Nodes}
\label{t10}
\begin{tabular}{|c|c|c|c|c|c|}
\hline
Pause  & \multicolumn{5}{c|}{Packets} \\
\hline
Time(s) & DTN   & Overall & Received & Send & Drop \\
\hline
10 & 0 & 611963 & 203710 & 12 & 472\\
20 & 3050 & 7022 & 3859 & 3102 & 32\\
30 & 7951 & 16195 & 8197 & 7973 & 16\\
40 & 8119 & 16282 & 8129 & 8129 & 22\\
50 & 7703 & 16202 & 8435 & 7724 & 25\\
60 & 8001 & 16257 & 8203 & 8017 & 28\\
70 & 8157 & 16316 & 8159 & 8157 & 0\\
80 & 7690 & 15453 & 7756 & 7694 & 1\\
90 & 7860 & 15756 & 7889 & 7862 & 3\\
100 & 8312 & 16629 & 8313 & 8313 & 3\\
\hline
\end{tabular}
\end{table}

\begin{table}[t]
\centering
\caption{Parameters for Scenario 5}
\label{t11}
\begin{tabular}{|c|c|c|}
\hline
Sr. NO. & Parameter & Value \\
\hline
1 & Number of Node & 150 \\
2& Total FTP Connection & 10 \\
\hline
3 & \multicolumn{2}{|c|}{ FTP(0) in Cluster 1} \\
 & Start time: 10s &  End time:80s\\
 & Source Node: Wired Node(0) & Dest Node: Mobile Node(0) \\
$ \vdots $ & $ \cdots $ & $ \cdots $ \\
\hline
10 & \multicolumn{2}{|c|}{ FTP(14) in Cluster 1} \\
 & Start time: 10s &  End time:90s\\
 & Source Node: Wired Node(0) & Dest Node: Mobile Node(14) \\
\hline
$ \vdots $ & $ \cdots $ & $ \cdots $ \\
\hline
11 & \multicolumn{2}{|c|}{ FTP(30) in Cluster 2} \\
 & Start time: 20s &  End time:90s\\
 & Source Node: Wired Node(1) & Dest Node: Mobile Node(30) \\
\hline
$ \vdots $ & $ \cdots $ & $ \cdots $ \\
\hline
12 & \multicolumn{2}{|c|}{ FTP(54) in Cluster 2} \\
 & Start time: 20s &  End time:90s\\
 & Source Node: Wired Node(1) & Dest Node: Mobile Node(54) \\
\hline
\end{tabular}
\end{table}

\begin{table}[t]
\caption{Simulation Results of Scenario 5:\emph{150} Mobile Nodes}
\label{t12}
\centering
\begin{tabular}{|c|c|c|c|c|c|}
\hline
Pause  & \multicolumn{5}{c|}{Packets} \\
\hline
Time(s) & DTN   & Overall & Received & Send & Drop \\
\hline
10 & 0 & 626544 & 208456 & 12 & 683\\
20 & 3114 & 8044 & 4770 & 3189 & 44\\
30 & 8094 & 16153 & 8597 & 8123 & 20\\
40 & 8210 & 16482 & 8222 & 8224 & 34\\
50 & 7804 & 17314 & 9406 & 7834 & 37\\
60 & 8127 & 16786 & 8583 & 8150 & 40\\
70 & 8108 & 16213 & 8105 & 8108 & 0\\
80 & 8093 & 16188 & 8095 & 8093 & 0\\
90 & 7950 & 15900 & 7950 & 7950 & 0\\
100 & 8513 & 17194 & 8671 & 8518 & 2\\
\hline
\end{tabular}
\end{table}

\begin{table}[t]
\centering
\caption{Parameters for Scenario 5}
\begin{tabular}{|c|c|c|}
\hline
Sr. NO. & Parameter & Value \\
\hline
1 & Number of Node & 200 \\
2& Total FTP Connection & 14 \\
\hline
3 & \multicolumn{2}{|c|}{ FTP(0) in Cluster 1} \\
 & Start time: 10s &  End time:80s\\
 & Source Node: Wired Node(0) & Dest Node: Mobile Node(0) \\
$ \vdots $ & $ \cdots $ & $ \cdots $ \\
\hline
10 & \multicolumn{2}{|c|}{ FTP(18) in Cluster 1} \\
 & Start time: 10s &  End time:90s\\
 & Source Node: Wired Node(0) & Dest Node: Mobile Node(14) \\
\hline
11 & \multicolumn{2}{|c|}{ FTP(40) in Cluster 2} \\
 & Start time: 20s &  End time:90s\\
 & Source Node: Wired Node(1) & Dest Node: Mobile Node(40) \\
\hline
$ \vdots $ & $ \cdots $ & $ \cdots $ \\
\hline
14 & \multicolumn{2}{|c|}{ FTP(136) in Cluster 2} \\
 & Start time: 20s &  End time:90s\\
 & Source Node: Wired Node(1) & Dest Node: Mobile Node(136) \\
\hline
\end{tabular}
\label{t11}
\end{table}

\begin{table}[t!]
\caption{Simulation Results of Scenario 6:\emph{200} Mobile Nodes}
\label{t14}
\centering
\begin{tabular}{|c|c|c|c|c|c|}
\hline
Pause  & \multicolumn{5}{c|}{Packets} \\
\hline
Time(s) & DTN   & Overall & Received & Send & Drop \\
\hline
10 & 0 & 639122 & 212658 & 12 & 843\\
20 & 3484 & 8639 & 5032 & 3546 & 32\\
30 & 7884 & 16334 & 8403 & 7906 & 16\\
40 & 8214 & 16475 & 8227 & 8224 & 22\\
50 & 8063 & 18206 & 10072 & 8084 & 25\\
60 & 8358 & 17339 & 8928 & 8374 & 28\\
70 & 8199 & 16398 & 8199 & 8199 & 0\\
80 & 8201 & 16403 & 8202 & 8201 & 0\\
90 & 8139 & 16283 & 8144 & 8139 & 0\\
100 & 8477 & 16946 & 8469 & 8477 & 0\\
\hline
\end{tabular}
\end{table}

\section{Results and their Analysis} \label{res}
\begin{figure}[h]
\centering
\mbox{\includegraphics[width=3in,height=2in]{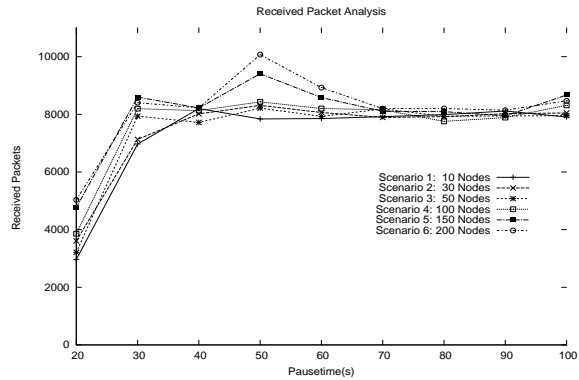}}
\caption{Packets Received in all Scenario}
\label{g1}
\end{figure}
In this section, the results gained are summarized graphically. Graph given in figure~\ref {g1} is plotted against Pausetime(s) verses number of received delay tolerant packets (DTN Packets) in all the Scenario considered in section \ref{sim}. 
\subsection{$\Delta$ and $\Theta$ analysis}
For this analysis, $ \Delta$ and $\Lambda_1$ are calculated. $ \Lambda_1 $  is defined as 
\begin{equation}
\Lambda_1   =  \frac {T  \* \Delta}{1000}  
\end{equation}
where $\Delta$ is given as 
\begin{equation} 
\Delta =  \frac{F_c}{N} 
\end{equation}

\begin{figure}[h]
\centering
\mbox{\includegraphics[width=3in,height=2in]{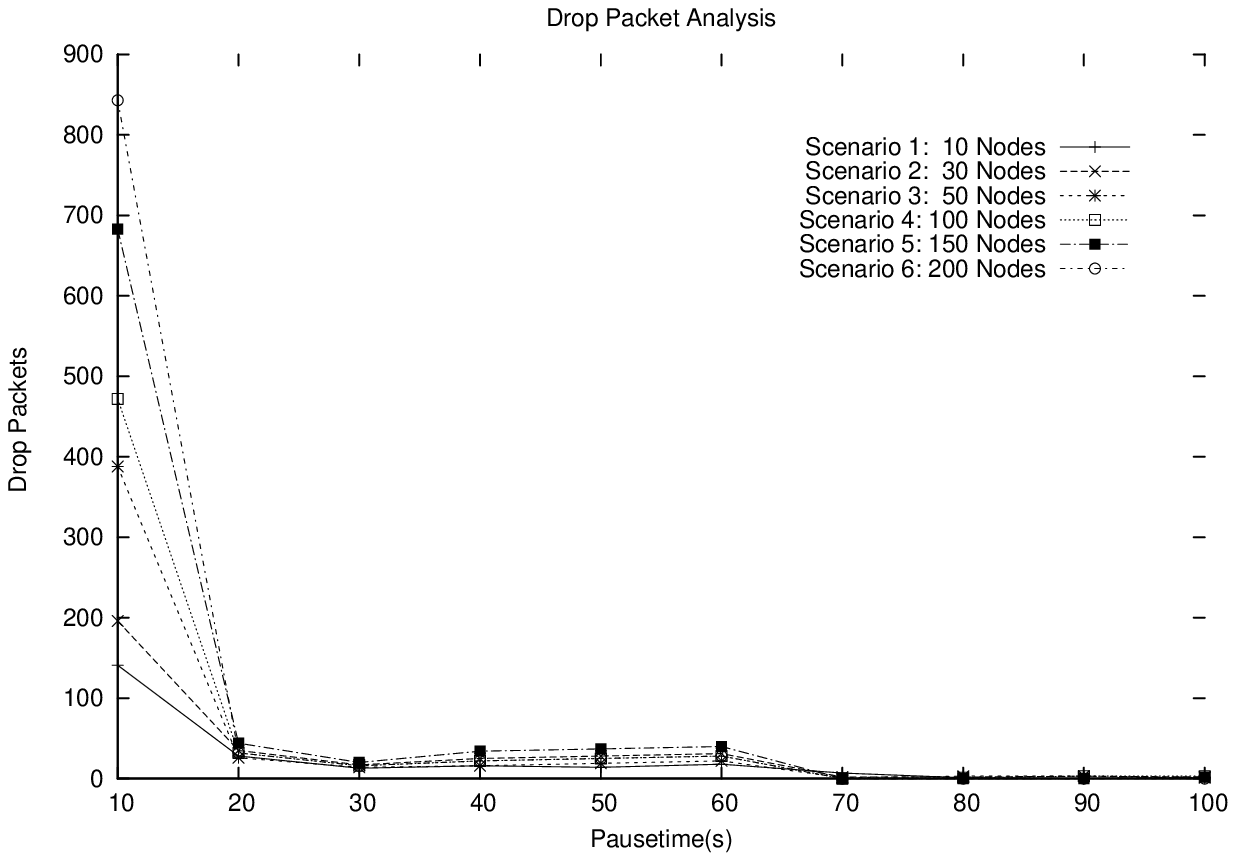}}
\caption{RTR and MAC layer packets dropped in the network}
\label{g2}
\end{figure}

$ \Lambda_1 $ is calculated for all delay tolerant network packets sent in the network for mentioned File Transfer Protocol connection according to scenario considered. 
\begin{equation}
T = \sum{_{1}^{10}} \; X(i,j)
\end{equation}
where X(i\,, j) represent row \emph{i} column \emph{j} intersection value where $ i= 1,2,\cdots, 10 $ and $ j=2 $ for each of the result table given in simulation section \ref{sim}.
From the table \ref{t15}, it is observed that the $\Lambda_1$ and $\Delta$  are decreases when the number of total FTP connection and number of Mobile node get increased. Another observation is the number of DTN packets sent are directly proportional to the number of FTP connection setup in the network. More the number of FTP connections, more will be the traffic and therefore more DTN packets will get sent from source node to the destination node. This can be clearly visualize from graph given in figure \ref{g2}. \\
We also calculate $ \Theta $ as 
\begin{equation}
\Theta_i =  \frac{PDTN_i  \Delta j }{100}
\end{equation}
where $PDTN_i$ represent the values of total Delay Tolerant Network Packets sent at $Pausetime_k$ for $ i=1,2,\cdots, 10 $, $ j=1,2, \cdots ,6$ \& $ k=10,20,\cdots ,100$. \\
Graph given in figure \ref{g2} is plotted against $\Theta$  vs. \emph{Pausetime(s)} for all scenario considered for simulation. From the graph we note that as the number of mobile node get increases, $\Theta$ get decreases and it becomes constant towards the average value of $\Theta$  for scenario considered as long as simulation continues. 
\begin{figure}[h]
\centering
\mbox{\includegraphics[width=3in,height=2in]{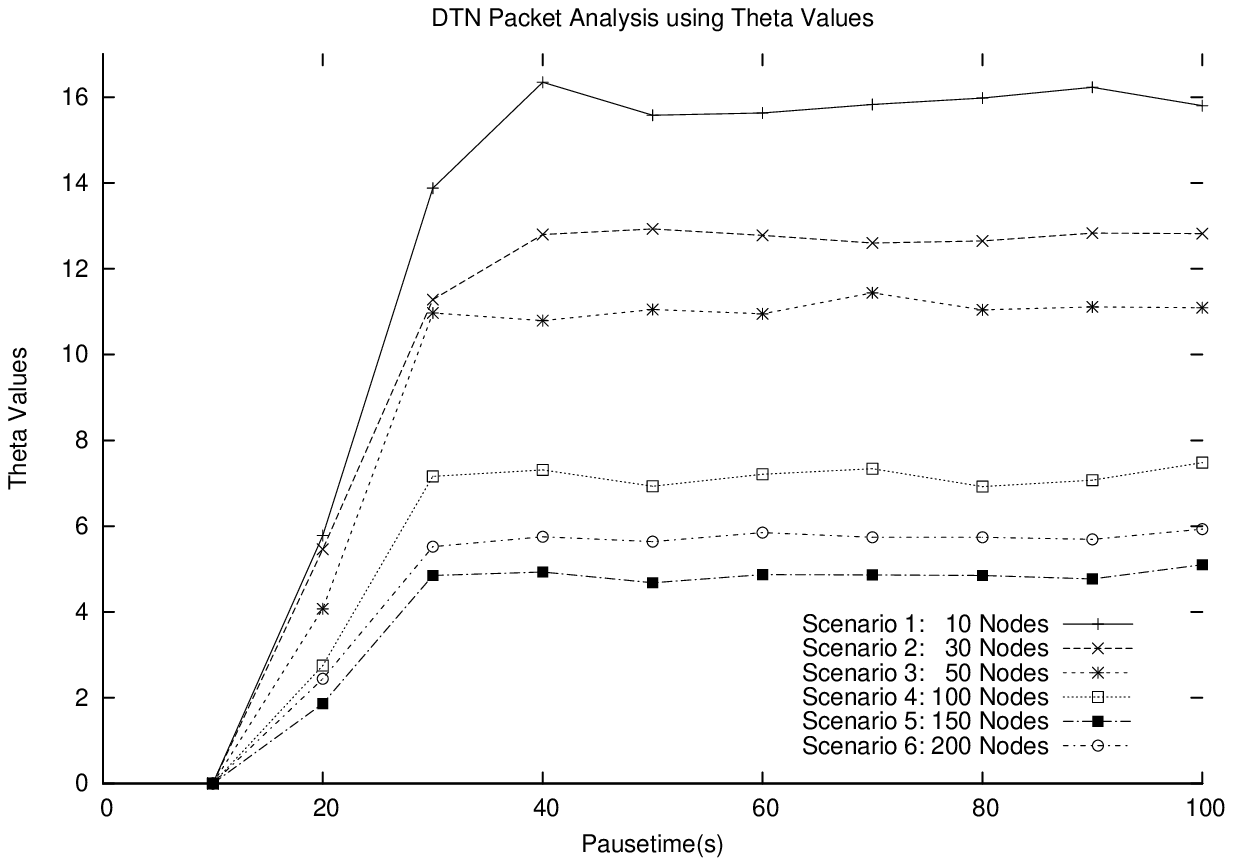}}
\caption{Various Theta values for all DTN Packets for all scenario}
\label{g5}
\end{figure}

\begin{table}[t]
\centering
\caption{$ \Delta \& \Lambda_1 $ for DTN Packets }
\begin{tabular}{|c|c|c|c|c|c|}
\hline
Sr.No. & Number of  & Number of  & Total DTN  & $ \Delta $ & $ \Lambda_1 $ \\
 &  Nodes(NN) &  FTP Connection $F_c$ &  Packets(T) && \\
\hline
1 & 10 & 02 & 65545 & 0.20 & 13.109 \\
2 & 30 & 05 & 66366 & 0.16 & 10.618\\
3 & 50 & 07 & 66104 & 0.14 & 9.254 \\
4 & 100 & 09 & 66843 & 0.09 & 6.015 \\
5 & 150 & 10 & 68013 & 0.06 & 4.080 \\
6 & 200 & 14 & 69019 & 0.07 & 4.831 \\
\hline
\end{tabular}
\label{t15}
\end{table}

\subsection{$\Lambda_2$ analysis}   
Another observation can be formulated by analyzing the factor $ \Lambda_2 $ as 
\begin{equation}
 \Lambda_2   = \frac {T \* \Delta}{10} 
\end{equation}
where $\Delta$ is same as given in equation (\emph{2}).
This factor $ \Lambda_2 $ is calculated for total number of packets which are dropped in the network. Since, DTN agent packets are sent over reliable data stream protocol called TCP, DTN packets are not dropped in the network and even if, it will be retransmitted due to properties of TCP Protocol. Therefore we analyze it for network layer (RTR) and Data Link Layer packets (MAC). We have, 
\begin{equation}
T = \sum{_{1}^{10}} \; X(i\,,j)
\end{equation}
where X(i\,, j) represent row \emph{i} column \emph{j} intersection value where $ i= 1,2,\cdots, 10 $ and $ j=6 $ for each of the result table.

From the table \ref{t16}, it is observed that the $\Lambda_2$ values increases when the Mobile node density and number of FTP connection increases. However, \emph{$\lambda_2 \le 7 $} and varies around \emph{6} for all scenario under consideration. This can be clearly foresee from graph given in figure ~ \ref{g3}. Also from the graph, Frequent packet drop occurs in the early stage of simulation where as it becomes almost constant as long as simulation continues.
\begin{table}[h]
\centering
\caption{$ \Delta \& \Lambda_2 $ for RTR and MAC Layer Packets dropped}
\begin{tabular}{|c|c|c|c|c|c|}
\hline
Sr.No. & Number of  & Number of  & Total RTR \& MAC  & $ \Delta $ & $ \Lambda_2 $ \\
 &  Nodes(NN) &  FTP Connection $F_c$ &  Packets(T) dropped && \\
\hline
1 & 10 & 02 & 239 & 0.20 & 4.78 \\
2 & 30 & 05 & 341 & 0.16 & 5.456\\
3 & 50 & 07 & 493 & 0.14 & 6.902 \\
4 & 100 & 09 & 602 & 0.09 & 5.418 \\
5 & 150 & 10 & 860 & 0.06 & 5.160 \\
6 & 200 & 14 & 966 & 0.07 & 6.762 \\
\hline
\end{tabular}
\label{t16}
\end{table}
   
\subsection{Throughput analysis}
\begin{figure}[h]
\centering
\mbox{\includegraphics[width=3in,height=2in]{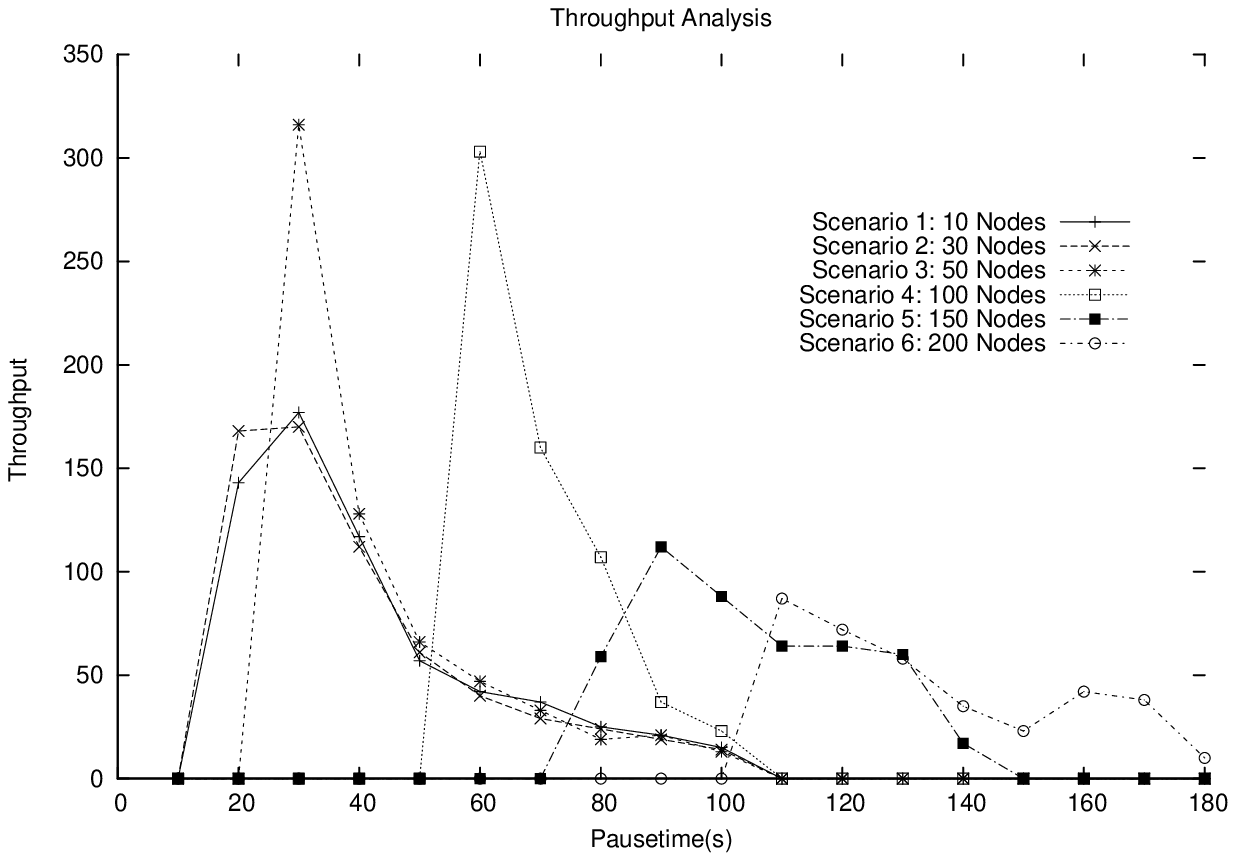}}
\caption{Throughput values for all DTN Packets}
\label{g3}
\end{figure}

Throughput is defined as the total number of bits sent per unit time. For throughput analysis, we plot the graph of Pausetime(s) against Number of DTN Packets received by the destination within the interval of \emph{0.5s} given in figure ~\ref {g3}. For throughput calculation , we consider packet size as 1540 bytes as the length of protocol data unit which includes 1460 bytes as data size received from higher layer and 80 bytes as header size. Throughput values are high for Scenario with mobile nodes \emph{50} and and mobile nodes \emph{100} nodes. Whereas it is low for rest of the Scenario with mobile nodes \emph{10,30} and also for the Scenario with mobile nodes \emph{150,200}. This clearly suggests the increasing number of nodes and FTP connection will decrease throughput after some threshold values. Moderate number of mobile nodes will give good throughput values.                  

\subsection{Routing Overhead}
\begin{figure}[t]
\centering
\mbox{\includegraphics[width=3in,height=3in]{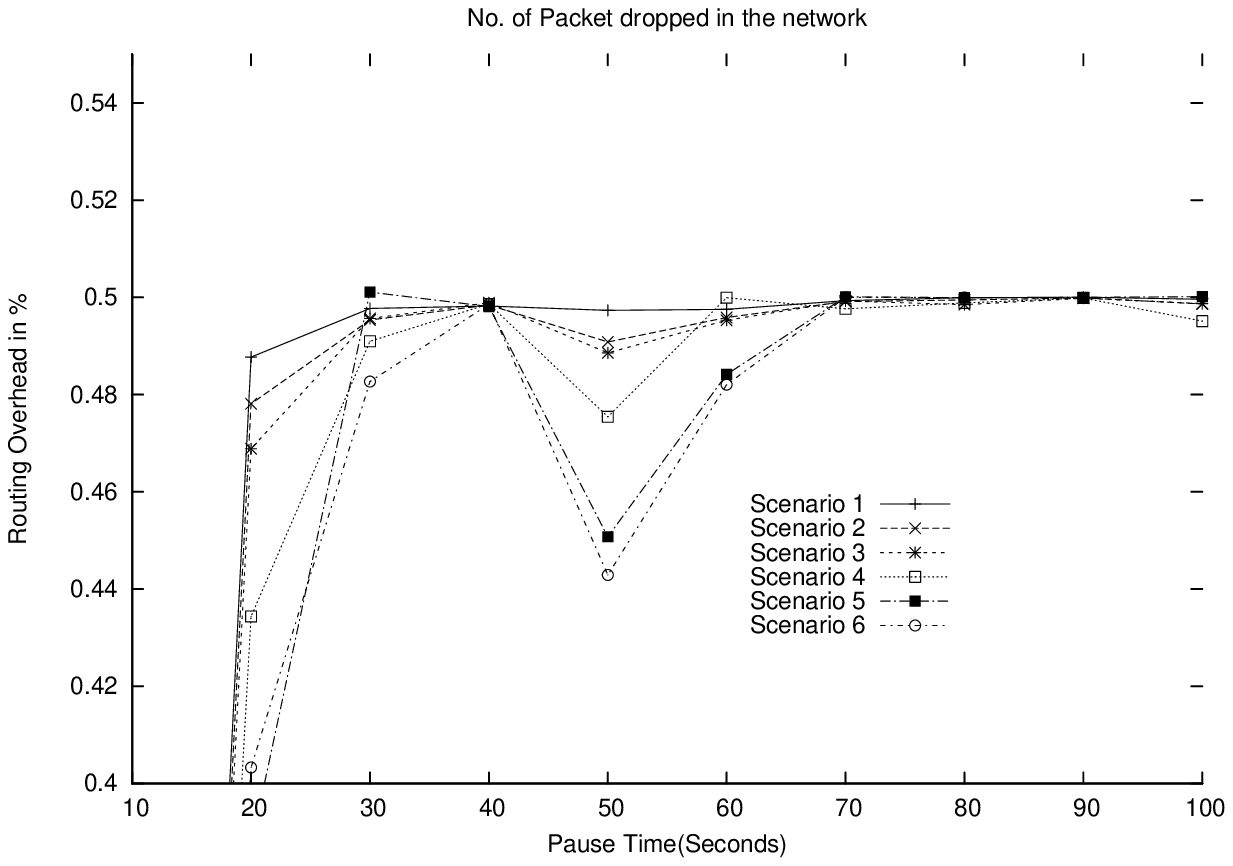}}
\caption{Routing Ovehead}
\label{g4}
\end{figure}

Routing overhead is defined as the ratio of total data packets sent and overall packets sent in the network. We plot \% of Routing Overhead verses Pausetime(s).
Another factor we considered Routing overhead. From the graph given in Figure ~\ref{g4}, it is viewed the Routing overhead is almost double compared to DTN Packets sent from source node to destination node for the scenario under consideration for every Pausetime. Usually routing overhead reflects the control packets sent in the network for finding route from source node towards destination. Therefore routing overhead increased as number of file transfer connection is increases along with number of mobile node. 

\section{Conclusion}\label{conl}
Delay tolerant network is opportunistic network. Research on delay tolerant network is continuously refined. The probability of transmission of message in delay tolerant network is high compared to adhoc network. In this paper, we evaluate the performance of delay tolerant network in a heterogeneous highly dense and dynamic mobile node environment. Heterogeneous nature of network is achieved using diversified network of nodes. This network we create with the help of Wired node, Mobile node as well as stationary gateway node called Base station node. Main focus of this study is to examine the effect of increasing number of mobile node and the traffic within these mobile nodes on the delay tolerant network. We create various scenarios by varying the number of Mobile nodes considered for simulation and the traffic routed between Wired node and Mobile node thorough Base station node. We analyze the results obtained by introducing three factors $\Delta$, $\Theta$ and $\lambda_1 $ and $\Lambda_2$. From the results we infer that increase in the number of mobile nodes does not affect throughput of the network. Throughput is high if suitable combination of mobile nodes and traffic is considered. Routing overhead has constant effect on the delay tolerant network traffic and almost 50 \% routing overhead is created for data traffic. Increase in the number of nodes and number of FTP connection rarely affect the overall performance of the network. In future, the performance of the delay tolerant network will be evaluated by varying Message storing capacity and transmission rate.

\bibliographystyle{amsalpha}
\bibliography{biblio}
\begin{figure}[h!]
\mbox{\includegraphics[width=4cm,height=4cm]{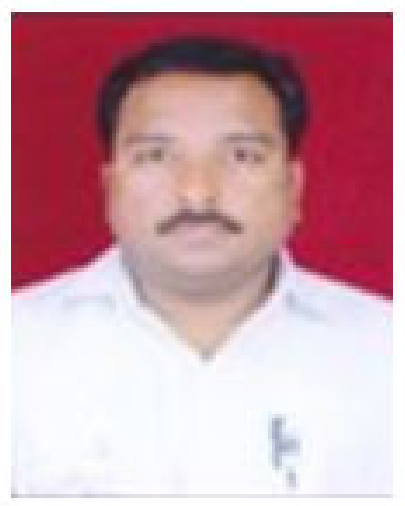}}
\\ \underline{Authors Biography} \\  $^1$ Prof. R.S.Mangrulkar is working as Associate Professor and head department of Computer Engineering, Bapurao Deshmukh College of Engineering, sevagram, Maharashtra, India. Currently he is persuing PhD  from SGBAU Amravati. He has total 12 years of teaching and research experience.  His area of interest is delay tolerant network, ad hoc network, cryptography.  He has published many research papers in national and international journal. He has delivered many guest lectures on Latex and its importance in research  writing and also conducted workshop on NS2. He is also serving as reviver of many International Journal.
\end{figure}
\begin{figure}[h!]
\mbox{\includegraphics[width=4cm,height=4cm]{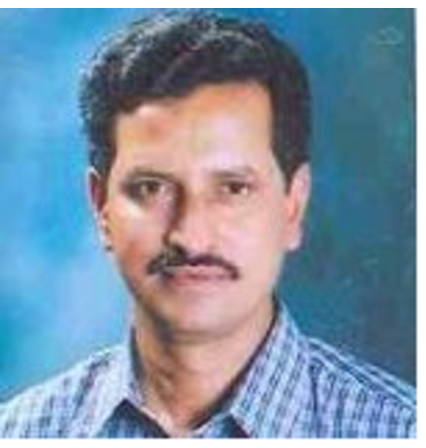}}
\\ \underline{Authors Biography} \\  $^2$ Dr. Mohammad Atique , is presently working as Associate Professor at Post Graduate Department of Computer Science, SGB Amravati University, Amravati since November 2007. He has more than 20 years of Teaching and administrative experience. He is Life Member of ISTE, New Delhi , IE Kolkata, Fellow IETE New Delhi and Sr. Life Member of CSI, Mumbai. His work gets recognized by many National and International journals. He is also reviver of reputed National and International journal. He is recognized as PhD supervisor for RTM Nagpur and SGBAU Amravati university. Currently 12 students are perusing PhD under him. His area of interest is Soft computing, Operating System, Ad hoc Network, Delay tolerant network, data mining and Machine Inteligent etc. He has chaired many sessions and also delivered keynote in National and International conference, STTPs in India. He also begs grant for his research work from various funding agency in India.                        

\end{figure}

\end{document}